# Large-scale Vortices in the Solar Photosphere: Statistical Properties and the Reliability of Local Correlation Tracking

Mashael Aldhafeeri[1,2], Suzana S. A. Silva[3], Istvan Ballai[1], Gary Verth[1], Haruhisa Iijima[4,5], Hideyuki Hotta[4], and Viktor Fedun[3]

[1] Plasma Dynamics Group, School of Mathematical and Physical Sciences, The University of Sheffield, Hicks Building, Hounsfield Road, Sheffield, S3 7RH, UK
[2] Department of Mathematics, College of Science, University of Hafr Al Batin, Hafr Al Batin, 39524, Saudi Arabia
[3] Plasma Dynamics Group, School of Electrical and Electronic Engineering, The University of Sheffield, Sir Frederick Mappin Building, Mappin Street, Sheffield, S1 3JD, UK
[4] Institute for Space-Earth Environmental Research, Nagoya University, Furocho, Chikusa-ku, Nagoya, Aichi 464-8601, Japan
[5] Institute for Advanced Research, Nagoya University, Furocho, Chikusa-ku, Nagoya, Aichi 464-8601, Japan
*Received 2026 July 9; revised 2026 September 7; accepted 2026 September 15; published 2026 October 19*

## Abstract

Vortices are observed throughout the solar atmosphere, spanning a wide range of spatial scales. Although small-scale vortices have been studied extensively, mesoscale vortices remain less explored, and it is still unclear whether they can be reliably recovered from observations. Mesogranular-scale motions are recovered from the R2D2 simulations by applying Gaussian spatial smoothing to the horizontal velocity field with full width at half maximum (FWHM) values of 2.5–4.0 Mm. To evaluate the accuracy of local correlation tracking (LCT) and its suitability for identifying mesogranular vortices, we apply LCT to synthetic intensity maps at $\tau = 1$ using the same spatial windows and temporal averaging windows of 10–60 minutes. Vortices are identified in both the reference and LCT-derived velocity fields using the $\Gamma$ method. For all tested parameter combinations, LCT underestimates the number of mesogranular-scale vortices. The best overall recovery is obtained for relatively small spatial smoothing scales (FWHM $\leqslant$3.5 Mm) and short averaging windows ($\leqslant$20 minutes), which provide the most reliable vorticity and lifetime statistics while maintaining strong agreement in vortex radius and area distributions. Our analysis shows that only a modest fraction of LCT-detected vortices correspond to vortices identified in the reference velocity field (3.54%–17.88%), depending on FWHM and temporal averaging. This indicates that vortex-driven dynamics inferred from LCT-detected vortices should be interpreted with caution. Our results suggest that applying the $\Gamma$ method to LCT-derived velocity fields can recover useful statistical properties of mesoscale vortical structures, especially their temporal trends and characteristic sizes.

## 1. Introduction

The solar atmosphere hosts several types of vortices, from rotational plasma flows in intergranular lanes to granular structures that follow vortical trajectories, producing larger-scale structures (K. Tziotziou et al. 2023). Of particular interest here are vortices associated with mesogranular and supergranular convection, which can have radii of several Mm and persist for tens of minutes to several hours (R. Attie et al. 2009; I. S. Requerey et al. 2018; A. C. Chian et al. 2019). For example, I. S. Requerey et al. (2018) studied a persistent photospheric vortex flow at a supergranular vertex and found that it consisted of three consecutive vortices, each with a lifetime of about 7 hr. More generally, mesogranular and supergranular vortex flows in the quiet Sun are thought to influence the distribution of magnetic fields (R. Attie et al. 2009; S. d. S. e. A. Silva et al. 2018; A. C. Chian et al. 2019).

In a velocity field, vortical motion is commonly associated with local rotation, with the vorticity defined as $\boldsymbol{\omega} = \nabla \times \boldsymbol{v}$. However, nonzero vorticity alone does not uniquely identify a vortex, since shear flows can also produce significant vorticity. Several methods have therefore been developed to identify coherent vortical structures, including IVD/LAVD (G. Haller et al. 2016; S. d. S. e. A. Silva et al. 2018; S. S. Silva et al. 2020), Rortex/Liutex (C. Liu et al. 2018; S. Tian et al. 2018; J. Canivete Cuissa & O. Steiner 2024), and the $\Gamma$ method (L. Graftieaux et al. 2001; I. Giagkiozis et al. 2018).

Photospheric vortex flows cannot be measured directly, but can instead be identified from horizontal velocity fields reconstructed from observations. Their identification and the physical properties derived from them therefore depend on the reliability of the method used to estimate the flow field. Several techniques have been developed to infer photospheric flows from observations, each based on different assumptions and input data. Local correlation tracking (LCT; L. J. November & G. W. Simon 1988), Fourier LCT (FLCT; G. H. Fisher & B. T. Welsch 2008), and coherent structure tracking (CST; M. Rieutord et al. 2007) infer optical flows from sequences of photospheric intensity images. Balltracking provides a faster alternative by following tracers advected by granulation patterns (H. Potts et al. 2004). Other methods infer flows from magnetic field measurements rather than intensity patterns. These include the minimum energy fit (MEF; D. Longcope 2004) and the differential affine velocity estimator (DAVE; P. W. Schuck 2006), which derive velocities from sequences of photospheric magnetograms. More recently developed, the deep-learning method DeepVel (A. A. Ramos et al. 2017) estimates horizontal pixel velocities from intensity images.

In this work, we focus on LCT, which has been widely used to study mesoscale solar dynamics and photospheric vortex flows (I. S. Requerey et al. 2018; S. d. S. e. A. Silva et al. 2018; A. C. Chian et al. 2019, 2023). In LCT, a Gaussian tracking window defines the local weighting of the cross-correlation and sets the effective spatial scale of the inferred flow. This is particularly relevant because the method tracks the evolution of intensity patterns rather than directly measuring plasma velocities. I. S. Requerey et al. (2018) applied LCT to Hinode/NFI observations using a Gaussian window with a full width at half maximum (FWHM) of ≈2.9 Mm, while B. Tremblay et al. (2018) evaluated LCT and FLCT using synthetic SDO/HMI intensitygrams with an effective FWHM of ≈2.5 Mm. Temporal averaging is commonly used to reduce short-timescale granular variability and noise (L. J. November & G. W. Simon 1988; M. Verma & C. Denker 2011), thereby improving sensitivity to longer-lived coherent motions at the mesogranular and supergranular scales. At the same time, stronger temporal smoothing can suppress short-lived or weak flow signatures.

Previous validation studies have shown that LCT and FLCT accuracy depend strongly on the adopted spatial and temporal smoothing. M. Rieutord et al. (2001) showed that granule motions do not reliably trace flows below spatial scales of approximately 2.5 Mm or on timescales shorter than approximately 0.5 hr. M. Verma et al. (2013) also showed that LCT tends to underestimate the ground-truth velocity field (GT), in particular at small scales. Furthermore, R. E. Louis et al. (2015) found that LCT accuracy is reduced in intergranular lanes and near granular edges. More recently, T. Kostić et al. (2026) found that FLCT also systematically underestimates flow velocities, with the degree of underestimation depending on the adopted tracking window size. Recent FLCT-based validation studies further highlighted that vortex identification in tracking-derived velocity fields can be highly sensitive to the choice of correlation kernel. However, these studies did not explicitly assess how errors in recovered flows can lead to incorrect vortex detections or distort the measured properties of the vortices.

B. Welsch et al. (2007) carried out a systematic comparison of several velocity-inversion techniques, including LCT and FLCT, using synthetic magnetograms and known velocity fields from an MHD simulation. Their study quantified how accurately the different methods reconstructed the underlying flows, and showed that substantial errors can remain depending on the adopted technique and temporal sampling. In contrast, the present study addresses a complementary question: whether mesogranular vortices identified from LCT-derived intensity motions correspond to vortices present in the underlying simulated photospheric velocity field, and how reliably their occurrence, radius, lifetime, vorticity, and spatial correspondence are recovered.

More recently, S. S. Turkay et al. (2026) explored vortex-level correspondence in greater detail, finding that, while vortex lifetime and area distributions may appear broadly similar, vortices detected in FLCT-derived velocity fields show only limited spatial correspondence with those identified in the underlying MHD velocity field. They therefore concluded that FLCT is unreliable for inferring photospheric intensity vortices in intergranular lanes.

In this paper, we use the high-resolution R2D2 radiative MHD simulation (H. Hotta & H. Iijima 2020) as a reference to evaluate how well LCT recovers photospheric mesogranular vortices. We compare vortices identified directly in the simulation velocity field with those detected in LCT-derived flows. In both cases, vortices are identified using the Γ criterion, which allows us to assess the recovery of vortex number density, radius, lifetime, and vorticity, as well as the dependence of these properties on the photospheric magnetic field strength. Section 2 describes the simulation data, the LCT setup, the vortex identification method, and the statistical measures used. Section 3 presents the comparison between vortex properties obtained from the GT simulation velocity field and the LCT-derived velocity fields. Section 4 discusses the main results and summarizes the conclusions.

## 2. Methodology

### 2.1. R2D2 Simulation

We use data from the R2D2 radiative MHD simulation of H. Hotta & H. Iijima (2020) for testing the recovery of photospheric vortices. The simulation models the emergence of a twisted force-free magnetic flux tube initially located at a depth of 35 Mm below the photosphere, followed by its interaction with convective motions. This simulation was selected because it evolves from quiet Sun conditions to an active region, thereby providing a robust test of LCT performance across different magnetic field configurations.

The R2D2 code solves the radiative MHD equations assuming an ideal plasma (i.e., without explicit physical resistivity). The computational domain spans $98 \times 98$ Mm$^2$ in the horizontal directions, with periodic boundary conditions and a uniform grid spacing of 96 km pixel$^{-1}$. Vertically, the simulation extends over a depth of 200 Mm on a nonuniform grid, and reaches up to about ∼700 km above the $\tau = 1$ surface in the lower photosphere. The initial force-free magnetic flux tube is placed in the middle of the domain and interacts self-consistently with the convective flows, leading to the emergence of the magnetic field and formation of solar pores.

The time series has a cadence of 120 s and spans a total simulated duration of 180 hr, providing approximately 37 hr before the onset of flux emergence and more than 100 hr of the decay phase after the peak magnetic flux reaches the photosphere. For our analysis, we use photospheric quantities at optical depth $\tau = 1$, namely the continuum intensity, the line-of-sight (LOS) magnetic field $B_z$, and the horizontal velocity components, defined as $v_h = (v_x, v_y, 0)$. We analyze the full horizontal domain ($1024 \times 1024$ pixels, $98 \times 98$ Mm$^2$) for 651 consecutive snapshots spanning $t = 26.7$ to 48.3 hr, corresponding to a total duration of ∼21.7 hr. For simplicity, we will consider the initial time to be $t = 0$ and the final time to be 21.6 hr after the beginning of our analysis. The continuum intensity map from the R2D2 simulation at optical depth $\tau = 1$ is shown in Figure 1 at $t = 15.7$ hr from the lower limit of our selected time interval, corresponding to $t = 42.4$ hr in the original simulation time.

### 2.2. Velocity Field Recovery with LCT

In the present study, LCT is used to estimate apparent horizontal motions from the evolution of intensity patterns in consecutive images. The method divides an image into local tracking windows and determines the displacement that maximizes the cross-correlation between corresponding windows in successive frames. Dividing this displacement by the time interval between frames yields an estimate of the two-dimensional local horizontal velocity.

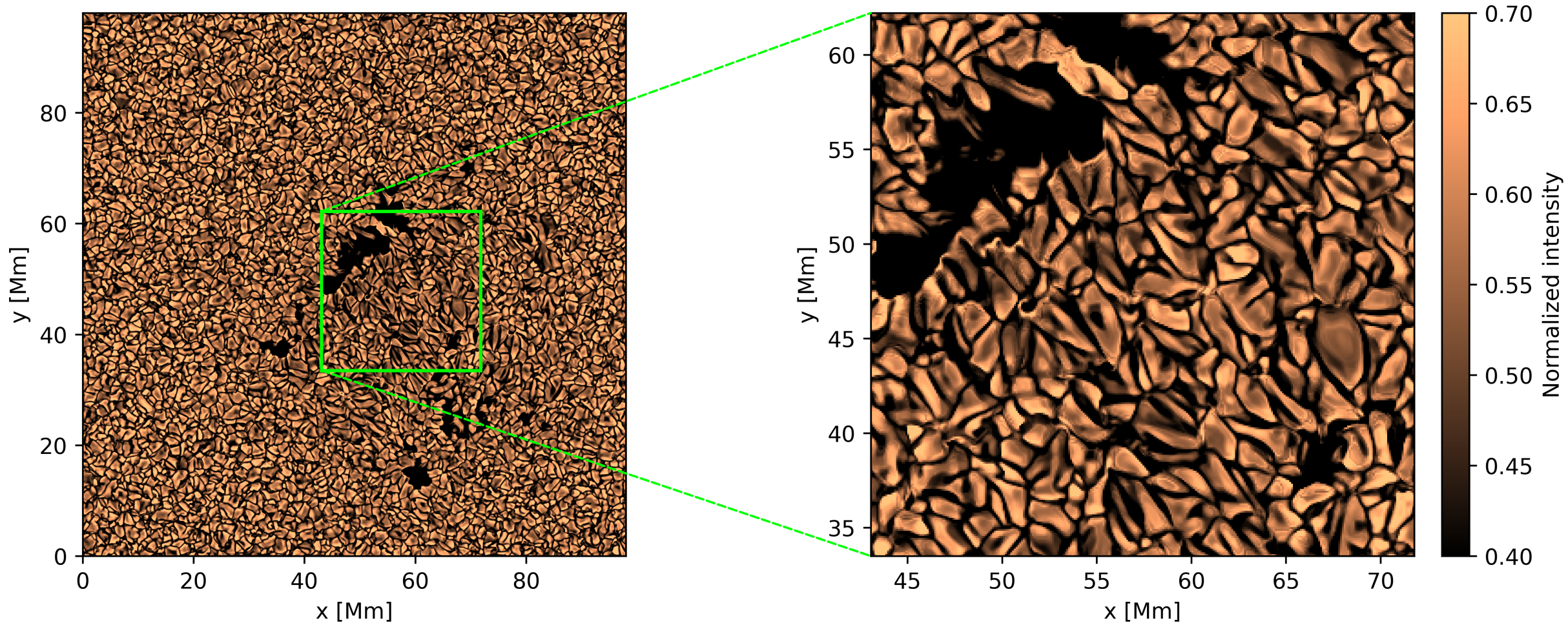

**Figure 1.** Continuum intensity at optical depth $\tau = 1$ from the R2D2 simulation at $t = 15.7$ hr from the lower limit of selected time interval. The left panel displays the full $98 \times 98$ $Mm^2$ field of view, while the right panel shows a zoom-in of the region outlined by the green box. Intensities are normalized by the maximum value in the field of view.

In this study, the R2D2 intensity images are tracked at the original simulation cadence of ($\Delta t = 120$ s). Gaussian tracking windows with FWHM values between 2.5 and 4 Mm are used, corresponding to approximately 26–42 pixels for the simulation plate scale of 96 km $pixel^{-1}$. We chose this range of FWHM to recover flows associated with mesogranular scales, which enables the detection of large-scale vortex flows.

The resulting LCT-derived velocity fields are temporally averaged over 10, 20, 30, and 60 minutes to suppress short-lived granular variability and enhance longer-lived mesoscale structures. At the same time, varying the averaging window allows us to test how increasing temporal smoothing affects the recovery of vortex properties. To ensure a meaningful comparison, the simulation horizontal velocity field is spatially smoothed using Gaussian kernels matched to the LCT tracking windows, so that the reference and reconstructed velocity fields represent comparable effective spatial scales. The resulting fields are then used as input for the vortex identification analysis.

### 2.3. Vortex Identification: Γ Method

For the automatic identification of vortices, we use the Γ method introduced by L. Graftieaux et al. (2001), which has been used by several earlier studies (e.g., I. Giagkiozis et al. 2018; S. d. S. e. A. Silva et al. 2018; Y. Yuan et al. 2023; L. McClure et al. 2026; S. S. Turkay et al. 2026). In the present study, the Γ method is applied in the same way to both the large-scale flows obtained from the numerical velocity fields and the LCT-derived velocity fields obtained from intensity, ensuring that the resulting vortex statistics are directly comparable.

The method defines two dimensionless scalar functions, $\Gamma_1$ and $\Gamma_2$, computed from the local horizontal velocity field in a neighborhood $S$ around each grid point $\boldsymbol{x}_p$. The first quantity, $\Gamma_1$, measures the average sense of rotation of the flow around $\boldsymbol{x}_p$, and it is defined in discrete form as

$$\Gamma_1(\boldsymbol{x}_p) = \frac{1}{|S|} \sum_{\boldsymbol{x}_m \in S} \frac{[(\boldsymbol{x}_m - \boldsymbol{x}_p) \times \boldsymbol{v}_m] \cdot \hat{\boldsymbol{z}}}{\|\boldsymbol{x}_m - \boldsymbol{x}_p\|_2 \, \|\boldsymbol{v}_m\|_2}, \tag{1}$$

where $S = \{\boldsymbol{x}_m\colon \|\boldsymbol{x}_m - \boldsymbol{x}_p\|_2 \leqslant R\}$ is a disk centered on $\boldsymbol{x}_p$.

We adopt $R = 15$ pixels (1.44 Mm), which sets the search window $S$ over which local rotational coherence is evaluated. In the above relation, $|S|$ is the number of points in $S$, $\| \cdot \|_2$ denotes the Euclidean norm, and $\hat{\boldsymbol{z}}$ is the unit vector normal to the flow plane. By construction, $|\Gamma_1| \leqslant 1$, and $|\Gamma_1|$ approaches unity at the center of an axisymmetric vortex.

Candidate vortex centers are identified as local maxima satisfying $|\Gamma_1| \geqslant 0.75$. This threshold is intentionally less restrictive than the value proposed by L. Graftieaux et al. (2001) for perfectly axisymmetric vortices. The lower threshold allows the detection of vortices with more irregular and less symmetric shapes, which are more representative of realistic solar flows. The second quantity, $\Gamma_2$, is used to delineate the vortex core. It is evaluated in the same search window $S$, but using velocities relative to the local mean velocity $\bar{\boldsymbol{v}}_p$, computed as the average of $\boldsymbol{v}_m$ within $S$:

$$\Gamma_2(\boldsymbol{x}_p) = \frac{1}{|S|} \sum_{\boldsymbol{x}_m \in S} \frac{[(\boldsymbol{x}_m - \boldsymbol{x}_p) \times (\boldsymbol{v}_m - \bar{\boldsymbol{v}}_p)] \cdot \hat{\boldsymbol{z}}}{\|\boldsymbol{x}_m - \boldsymbol{x}_p\|_2 \, \|\boldsymbol{v}_m - \bar{\boldsymbol{v}}_p\|_2}. \tag{2}$$

In the limit of small $S$, $\Gamma_2$ depends on the local balance between rotation and strain. The study by L. Graftieaux et al. (2001) showed that $|\Gamma_2| > 2/\pi$ corresponds to rotation-dominated flow, $|\Gamma_2| = 2/\pi$ to pure shear, and $|\Gamma_2| < 2/\pi$ to regions dominated by strain. In their application to finite kernels, the vortex inner core was defined using a threshold typically in the range 0.6–0.7, approximately $2/\pi$. In our study, vortices are identified in both the numerical velocity field and the velocity field reconstructed with the use of LCT using the same Γ-method implementation and identical parameter settings ($R$ and the thresholds defined above). This ensures that the resulting vortex statistics are directly comparable across the two datasets.

We also assess the sensitivity of the detected vortex statistics to the choice of center threshold $|\Gamma_1|$, using a

**Table 1**
Vortex Detection Statistics as a Function of the Center Threshold $|\Gamma_1|$, for a Representative Case (FWHM =3.5 Mm; 30 Minute Averaging)

| $\|\Gamma_1\|$ | $\Delta N_{GT}$ | $\Delta N_{LCT}$ | $N_{LCT}/N_{GT}$ |
|---|---|---|---|
| $\geqslant$0.70 | … | … | 0.513 |
| $\geqslant$0.75 | −31.5% | −34.7% | 0.489 |
| $\geqslant$0.80 | −56.0% | −57.5% | 0.495 |

**Note.** Percentage decreases are relative to the $|\Gamma_1| \geqslant 0.70$ baseline.

representative case (FWHM ≈3.5 Mm; 30 minute averaging) with the boundary threshold fixed at $|\Gamma_2| \geqslant 0.6362$. As $|\Gamma_1|$ increases from 0.70 to 0.80, the total number of detections falls substantially in both fields (Table 1). The detection efficiency is quantified using the ratio $N_{LCT}/N_{GT}$, where $N_{LCT}$ and $N_{GT}$ are the total vortex counts in the LCT-derived and numerical velocity fields, respectively. The ratio remains stable near 0.5 throughout, confirming that the LCT-derived field consistently recovers approximately half of the vortex detections found in the numerical velocity field, regardless of the chosen threshold.

### 2.4. Overlap-based Matching Criterion

To quantify the correspondence between vortices detected in LCT-derived velocity fields and those detected in the GT velocity fields, we applied an overlap-based criterion. For each vortex identified in the GT velocity field at time step $t$, candidate LCT detections were searched within the time interval $t \pm 1$, corresponding to ±2 minutes for the cadence used here. The following three criteria were then used to identify overlapping vortex detections:

1. The center of the nearest LCT-detected vortex must lie within two pixels of the center of the GT vortex.
2. The overlap fraction between the two $\Gamma_2$-defined vortex regions must be at least 0.3. Here, the vortex region refers to the set of pixels enclosed by the Γ-defined vortex boundary, and the overlap fraction is defined as the area of intersection divided by the area of union of the two regions.
3. The local mean vertical vorticities of the GT and LCT-derived velocity fields must have the same sign, ensuring that the two detections have the same sense of rotation.

If all three criteria were satisfied, the LCT detection was classified as overlapping with the corresponding GT vortex. If one or more criteria were not satisfied, the LCT detection was treated as having no corresponding vortex in the GT velocity field. This procedure was used to calculate the overlap percentage between the LCT-derived and GT vortex populations for each FWHM and temporal averaging combination.

## 3. Results

Figures 2 and 3 show examples of vortices identified in both fields at $t$ ≈5 hr 22 minutes after flux emergence. In both figures, the marked points indicate vortex centers identified with the Γ method, while the closed contours show the corresponding vortex boundaries. The squares in the main panels mark the regions of interest (ROIs) displayed in the enlarged panels.

The right panels (a)–(g) in Figure 2 show ROIs selected from vortices detected in the GT velocity field. Each ROI compares the same location in the GT velocity field, shown in the upper subpanel outlined in blue, with the corresponding LCT-derived velocity field, shown in the lower subpanel outlined in red. Figure 3 shows the equivalent comparison for vortices detected in the LCT-derived field. In this case, each ROI shows the LCT-derived field in the upper subpanel outlined in red and the corresponding GT velocity field in the lower subpanel outlined in blue.

In both Figures 2 and 3, the region shown in panel (d) is the only case where the GT and LCT-derived fields exhibit a closely matching rotational pattern, with nearly coincident vortex centers and the same sense of rotation. This indicates that applying the Γ method to LCT-derived velocity fields can recover individual vortices in favorable cases. However, most other detections do not exhibit a clear one-to-one correspondence. In these cases, the LCT-derived velocity field often shows curved or saddle-like streamlines, but the reconstructed flow pattern is distorted relative to the corresponding GT vortex. This limited correspondence is likely related to the way LCT estimates the flow. Rather than measuring the plasma motion directly, LCT follows changes in the intensity pattern within a finite tracking window. The evolution of the granulation, including changes in shape, splitting, or disappearance of features, can therefore influence the inferred motion. This can shift or distort the recovered velocity pattern and move the apparent vortex center away from its GT position. In some cases, these distortions may also produce a rotational pattern that satisfies the vortex detection criterion even when no corresponding GT vortex is present, resulting in a false-positive detection. The vortex structure is one of the most sensitive to the quality of velocity recovery, because it depends critically on correctly recovering the direction and local organization of the velocity vectors. Even when the overall flow amplitude is reasonably reproduced, small directional errors can erase the rotational pattern and therefore displace centers or recover instead of the vortex pattern a saddle or just a curved streamline, as observed in Figures 2 and 3

To quantify this visual impression, we use the overlap-based matching criterion described in Section 2.4. Accordingly, the fraction of LCT detections overlapping with GT vortices ranges from 3.54% to 17.88%, depending on the FWHM and the temporal averaging window. This confirms that, while individual vortex identification is possible under suitable parameter settings, the majority of LCT detections do not correspond to a unique counterpart in the GT velocity field.

### 3.1. Time-dependent Comparison

We first examine whether applying the Γ method to LCT-derived velocity fields reproduces the temporal evolution of the vortex population in the simulation. In this part of the analysis, we compare the time variation of vortex occurrence, recovery fraction, median radius, lifetime, and vorticity. To characterize the time-dependent occurrence of vortices, we compare the per-frame counts in the GT and LCT-derived velocity fields, denoted by $N_{GT}(t)$ and $N_{LCT}(t)$, respectively. For each time step, we define the relative count ratio

$$R(t) = \frac{N_{LCT}(t)}{N_{GT}(t)}, \quad (3)$$

which measures whether the velocity field derived using LCT yields fewer or more vortex detections than the GT field at the same time. Time steps with $N_{\rm GT}(t) = 0$ are excluded from statistics based on $R(t)$. To assess the cumulative detection performance, we also compute the integrated counts, $N_{\rm GT} = \sum_t N_{\rm GT}(t)$ and $N_{\rm LCT} = \sum_t N_{\rm LCT}(t)$, together with the total count ratio $R_{\rm tot} = N_{\rm LCT}/N_{\rm GT}$. The temporal mean, $\langle R \rangle$, and standard deviation, $\sigma_R$, are then computed over all valid frames as

$$\langle R \rangle = \frac{1}{N_f}\sum_{i=1}^{N_f} R_i, \quad \sigma_R = \left[\frac{1}{N_f - 1}\sum_{i=1}^{N_f}(R_i - \langle R \rangle)^2\right]^{1/2}, \quad (4)$$

where $N_f$ is the number of valid frames and $R_i$ is the ratio of frame $i$. These quantities are summarized in Table 1 for all tested combinations of spatial FWHM and temporal averaging window.

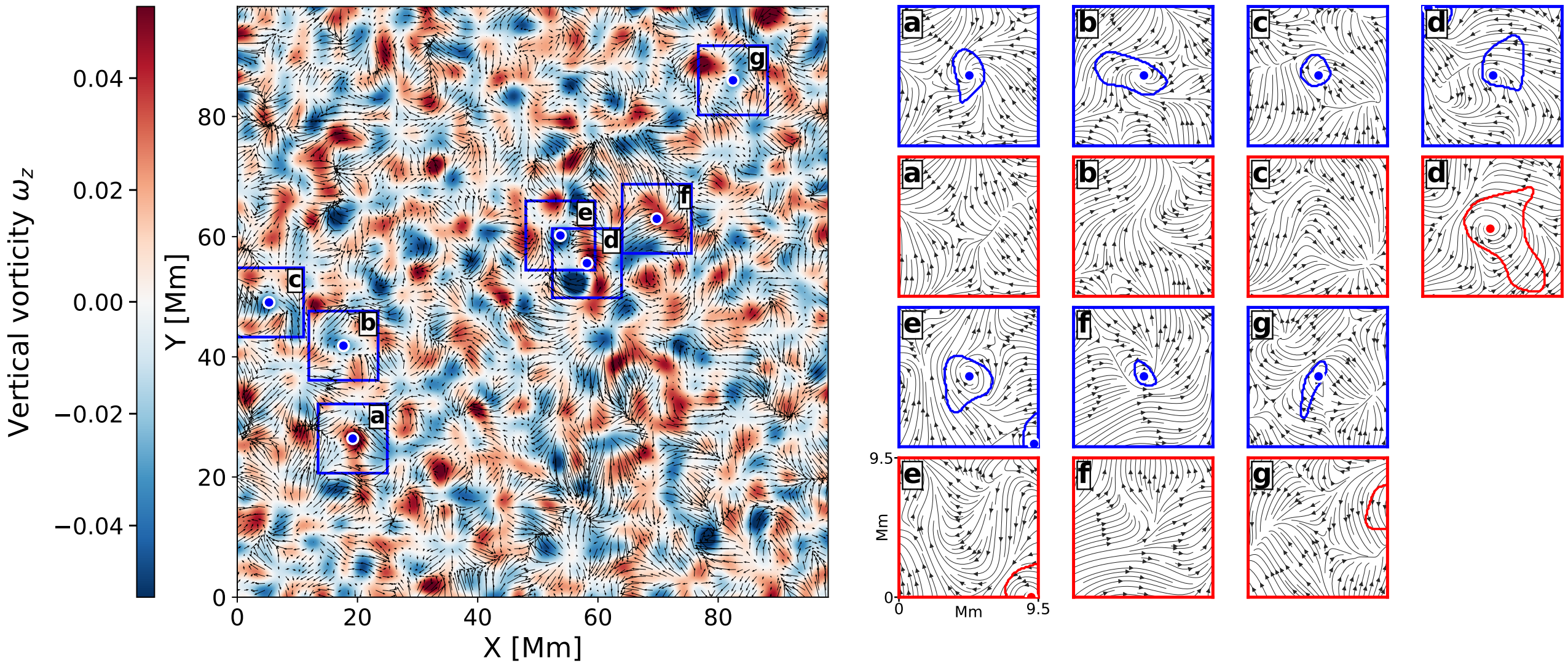

**Figure 2.** Comparison of vortices detected in the GT and LCT-derived velocity fields at $t \approx 5$ hr 22 minutes after flux emergence. Left panel: vertical vorticity, $\omega_z$, of the simulated velocity field after Gaussian spatial filtering with FWHM ≈3.5 Mm, with velocity vectors overplotted. Blue markers denote vortex centers identified using the Γ method. Right panels: enlarged views of the regions marked by the blue boxes in the left panel. For each case (a)–(g), the upper subpanel, outlined in blue, shows streamlines of the filtered simulation velocity field, while the lower subpanel, outlined in red, shows the corresponding LCT-derived velocity field at the same location. Blue and red contours indicate the vortex boundaries in the GT and LCT-derived velocity fields, respectively, and the filled markers denote the detected vortex centers.

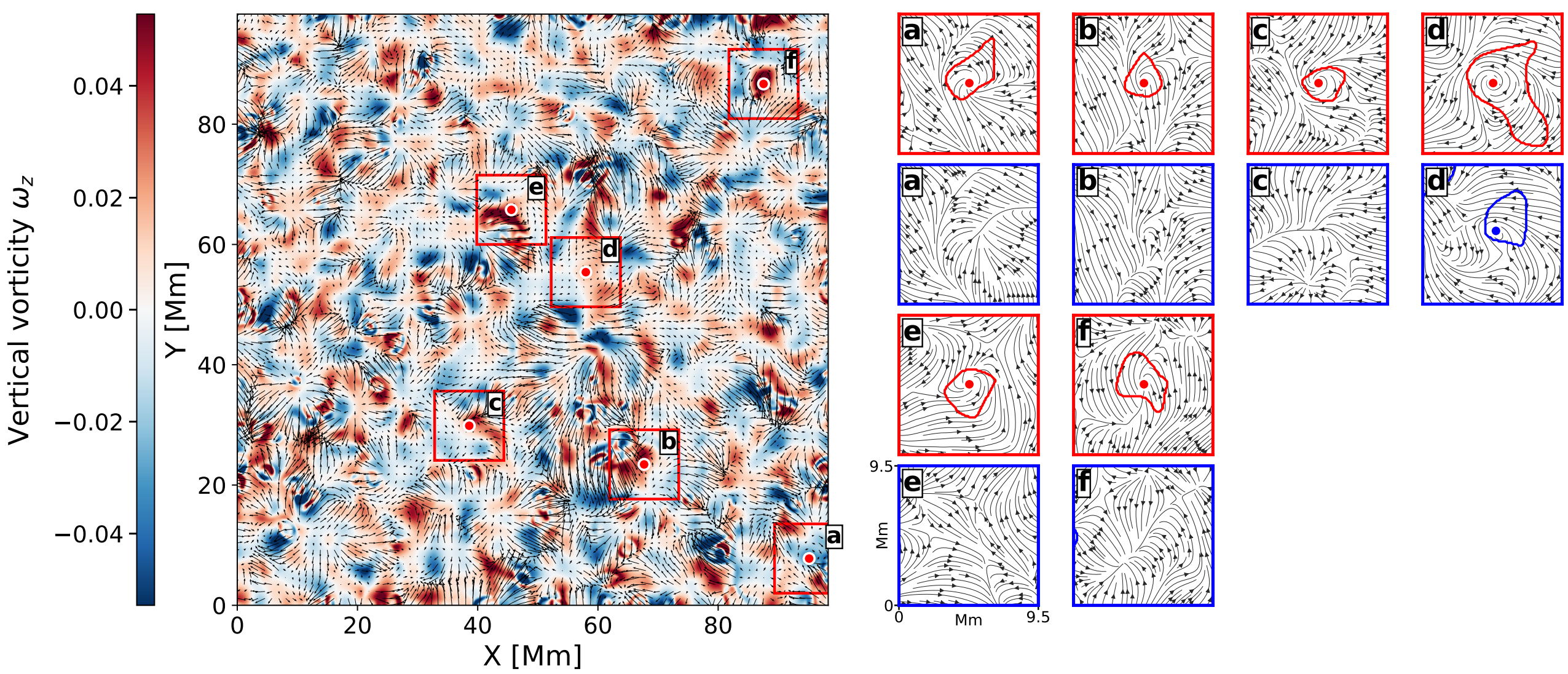

**Figure 3.** Same as Figure 2, but for regions selected from vortices detected in the LCT-derived velocity field. Left panel: vertical vorticity, $\omega_z$, of the LCT-derived horizontal velocity field over the full domain, with velocity vectors overplotted. Red markers denote vortex centers, and red boxes indicate the regions shown in the enlarged panels on the right. Right panels: enlarged views of the selected regions. For each case (a)–(g), the upper subpanel, outlined in red, shows streamlines of the LCT-derived velocity field, while the lower subpanel, outlined in blue, shows the corresponding GT velocity field at the same location.

Figure 4 shows the time variation of the relative count ratio as in Equation (3), for FWHM ≈3.5 Mm and temporal averaging windows of 10, 20, 30, and 60 minutes. This ratio quantifies how the number of vortices detected in the LCT-derived velocity field compares with the number detected in the GT velocity field. For the 10 minute window, the LCT-based counts can occasionally match or exceed the GT counts, but the ratio is highly variable. As the temporal averaging window increases, both the mean ratio $\langle R \rangle$ and its standard deviation $\sigma_R$ decrease. with $\langle R \rangle \approx 1.12$, 0.69, 0.56, and 0.38 for the 10, 20, 30, and 60 minute windows, respectively. The corresponding standard deviations are 0.70, 0.49, 0.43, and 0.40. This shows that stronger temporal smoothing progressively suppresses vortex detections in the velocity fields obtained using LCT.

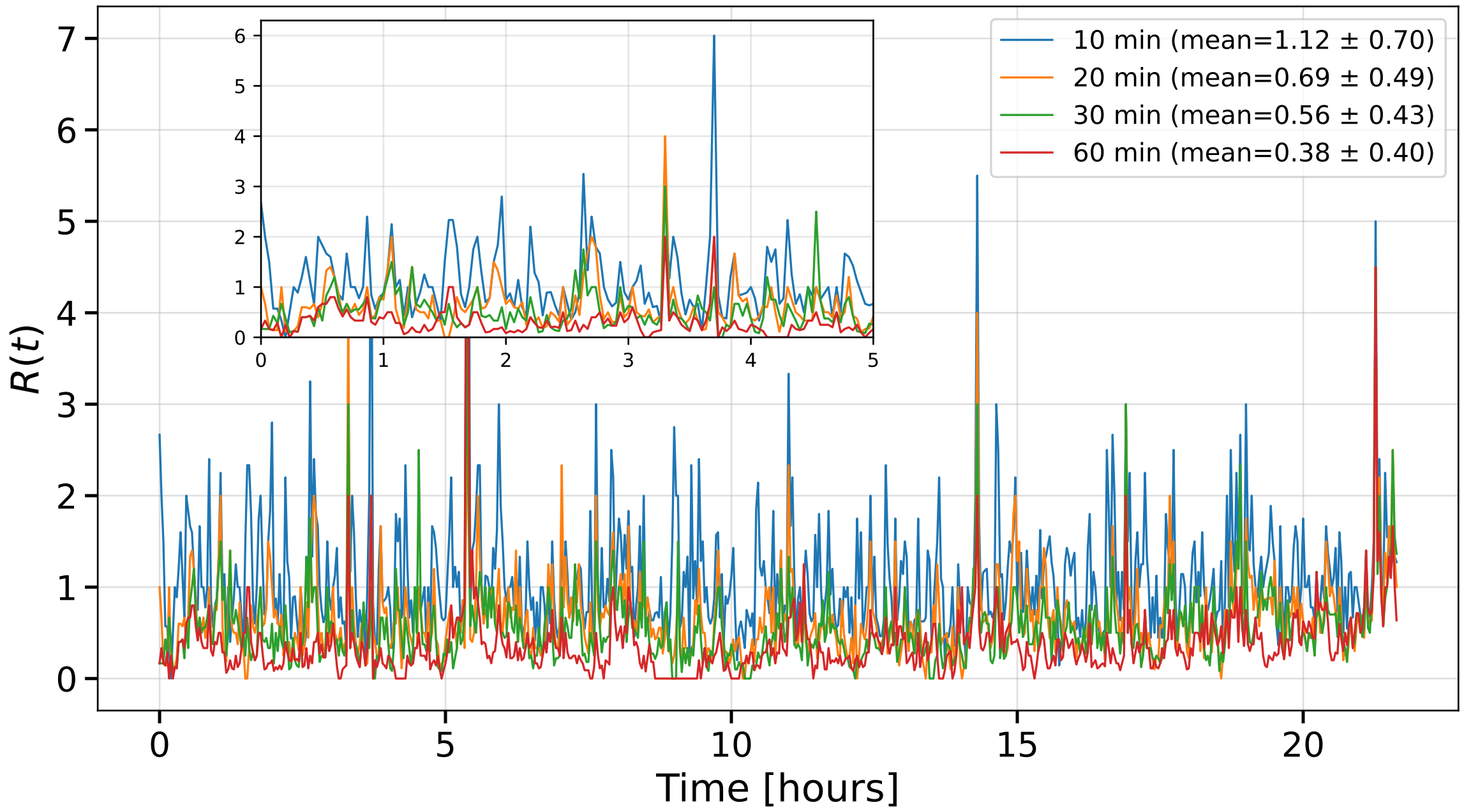

**Figure 4.** Ratio $R(t)$ for the representative spatial smoothing scale FWHM ≈3.5 Mm and temporal averaging windows of 10, 20, 30, and 60 minutes. The legend lists $\langle R \rangle$ and $\sigma_R$ for each averaging window. The inset shows the interval from 1 to 5 hr.

To quantify whether LCT reproduces the temporal variability of vortex occurrence, we compute Spearman's rank correlation coefficient $\rho$ between vortex-count time series obtained from the GT, $N_{\rm GT}(t)$, and LCT-derived velocity fields, $N_{\rm LCT}(t)$. Spearman's $\rho$ measures the strength of the monotonic relationship between the two time series and is computed from their ranked values. The vortex-count time series from both methods were smoothed using a centered moving average with the same duration as the temporal averaging window applied to the LCT velocity field. This ensures that the correlation is evaluated on the same temporal scales in both datasets, since temporal averaging of the LCT velocity field suppresses short-timescale fluctuations and retains only longer-timescale variations in vortex occurrence. Given the simulation cadence of 2 minutes, the 10, 20, 30, and 60 minute averaging windows correspond to moving average lengths of 5, 10, 15, and 30 frames, respectively. Spearman's coefficient, therefore, quantifies how well LCT reproduces the temporal evolution of the vortex counts on the timescales retained by each averaging window.

Since the paired time series are finite and autocorrelated (after temporal averaging), we estimate the uncertainty in $\rho$ using a circular block bootstrap for dependent data (D. N. Politis & J. P. Romano 1994). The paired time series are resampled in contiguous blocks of length $L$, where $L$ is defined in frames and chosen to match the temporal averaging window. With a cadence of 2 minutes, the 10, 20, 30, and 60 minute averaging windows correspond to $L = 5$, 10, 15, and 30 frames, respectively. For each bootstrap resample, we recompute $\rho$ and define the 95% confidence interval from the empirical 2.5th and 97.5th percentiles of the resulting bootstrap distribution. Correlations whose 95% confidence intervals include zero are regarded as statistically inconclusive. As shown in Table 1, the correlations between the numbers of GT and LCT-detected vortices range from weak to moderate. The strongest agreement is obtained for FWHM ≈3.5 Mm with a 30 minute averaging window, for which $\rho \simeq 0.53$.

Figure 5 shows the temporally smoothed vortex counts in the GT and LCT-derived velocity fields for the same FWHM ≈3.5 Mm case. The counts are smoothed using centered moving averages with windows of 10, 20, 30, and 60 minutes, matching the corresponding LCT temporal averaging windows. The domain-averaged magnetic field magnitude, $\langle |\boldsymbol{B}| \rangle$, is also shown on the right-hand axis and is smoothed using the same window in each panel. The LCT-derived vortex counts increase during the more strongly magnetized phase. However, the counts based on the LCT approach remain systematically below the GT counts, especially for the 30 and 60 minute averaging windows. To determine whether this recovery depends specifically on the local magnetic environment of individual vortices, we also measured the median total magnetic field strength within each defined GT vortex core and compared vortices that were successfully matched to LCT detections with those that were not. The median core-field strengths of the matched and unmatched populations were very similar for all four temporal averaging windows. In addition, the fraction of GT vortices recovered by LCT showed no systematic monotonic variation across quartiles of local core-field strength. We therefore find no clear evidence that the local magnetic field strength at the vortex core is a primary factor controlling LCT detection accuracy.

### *3.1.1. Radius Consistency*

We next test whether LCT reproduces the characteristic size of the detected vortices. Vortex size is quantified using the mean boundary radius, defined as the average distance from the vortex center to the vortex boundary. To reduce sensitivity to outliers and frame-to-frame fluctuations, we use the per-frame median radius as a representative measure. Figure 6 shows the temporal evolution of the median vortex radius for the representative spatial scale FWHM ≈3.5 Mm and temporal averaging windows of 10, 20, 30, and 60 minutes. The weak-and strong-field intervals are indicated by background shading and are defined using the domain-averaged total magnetic field strength. The threshold was chosen as $\langle|\boldsymbol{B}|\rangle = 80$ G, which occurs at approximately $t \simeq 15$ hr and corresponds approximately to the transition from the slowly varying pre-emergence phase to the more rapidly increasing magnetic field phase. Frames with $\langle|\boldsymbol{B}|\rangle < 80$ G are classified as weak field, while frames with $\langle|\boldsymbol{B}|\rangle \geqslant 80$ G are classified as strong field.

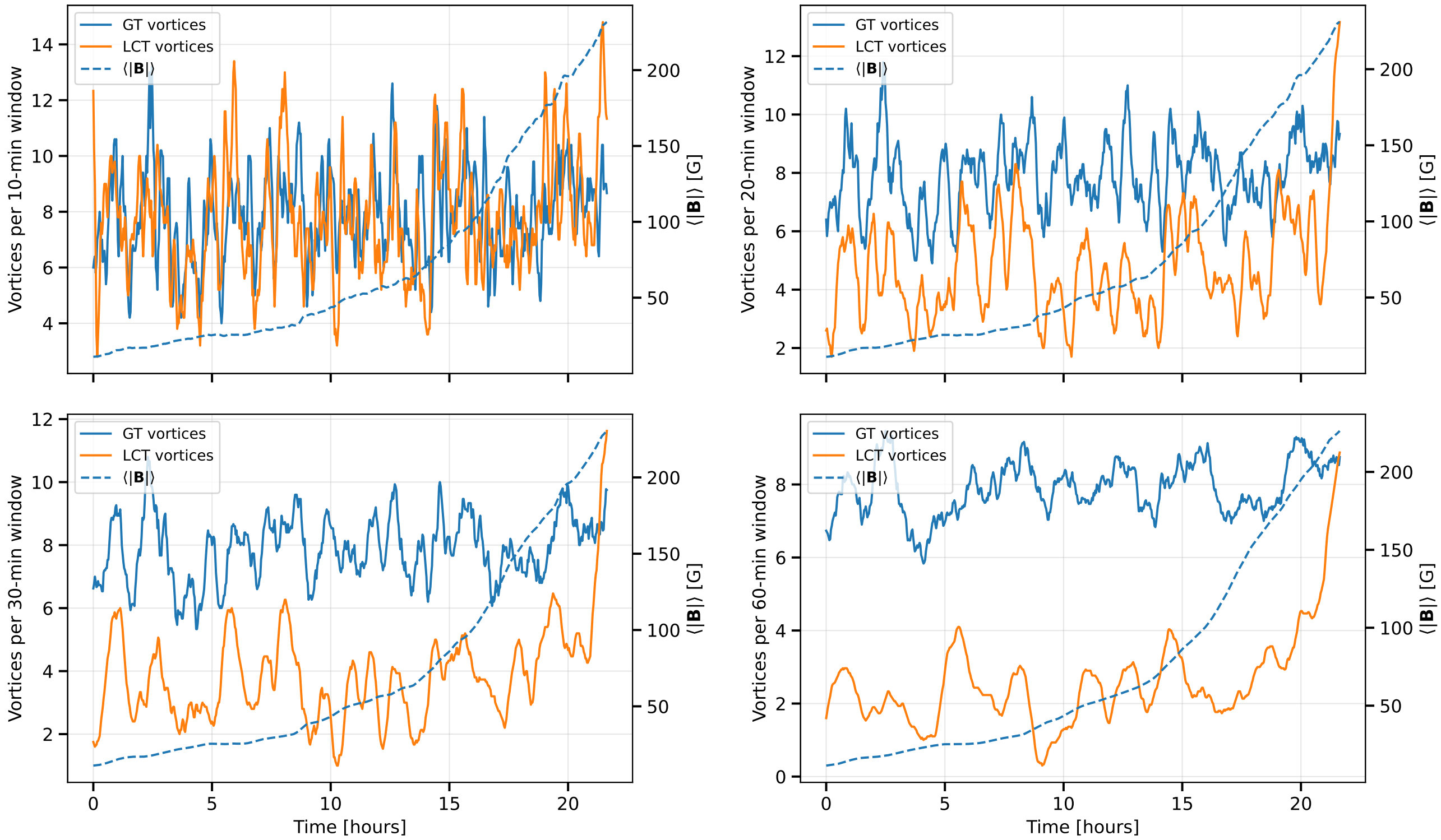

**Figure 5.** Time series of vortex detections identified with the Γ method in the GT velocity field (blue) and in the LCT-derived velocity field (orange), using FWHM ≈3.5 Mm. Panels show temporal averaging windows of 10, 20, 30, and 60 minutes. For each averaging time, the vortex-count curves show centered moving averages of the per-frame number of detections, using the same duration as the corresponding temporal window. The dashed curve, shown on the right-hand axis, gives the domain-averaged total magnetic field strength, $\langle|\boldsymbol{B}|\rangle$, smoothed with a centered moving average of the same duration.

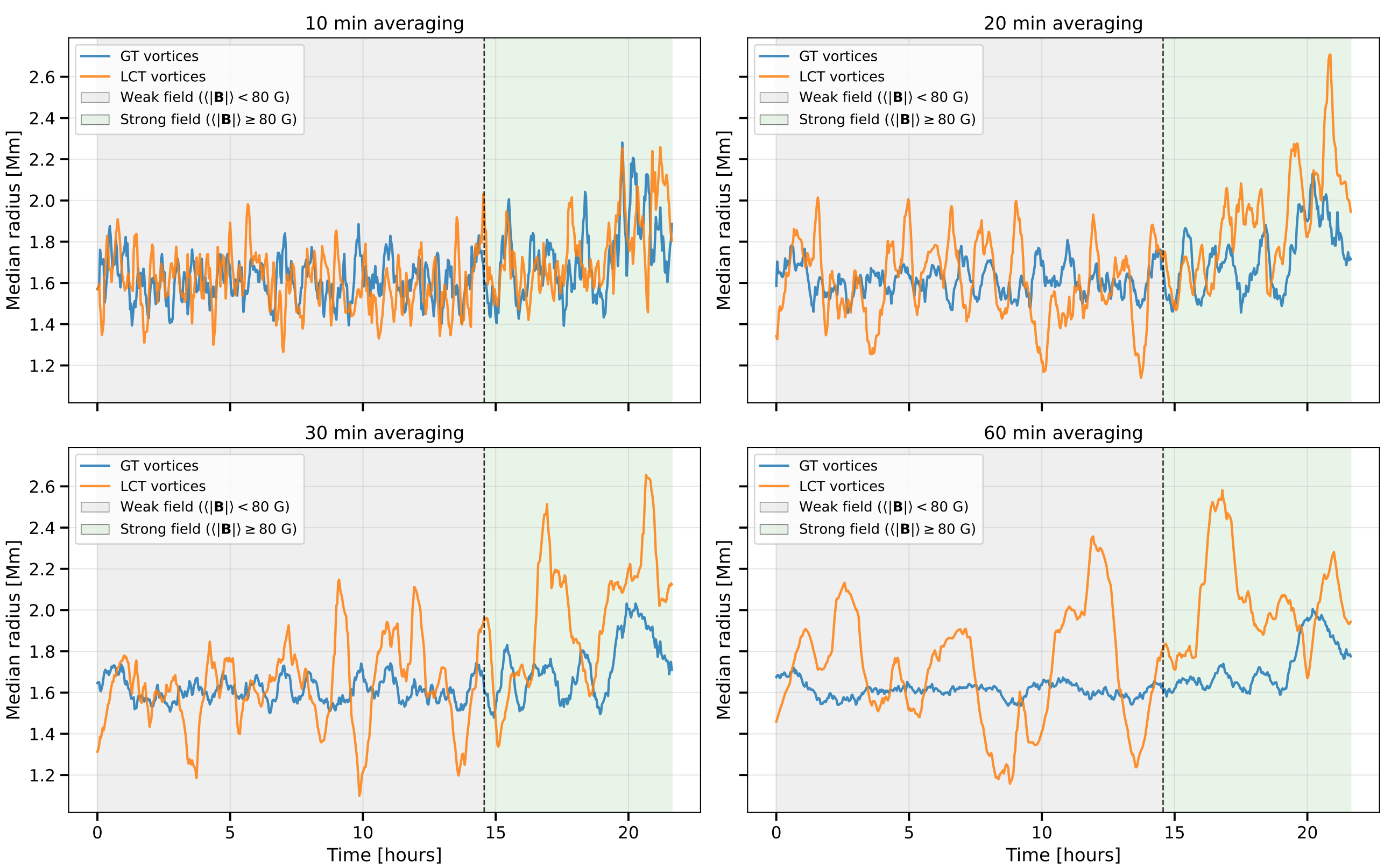

**Figure 6.** Temporal evolution of the median vortex radius for FWHM ≈3.5 Mm in the GT and LCT-derived velocity fields, shown for temporal averaging windows of 10, 20, 30, and 60 minutes. The curves are smoothed using moving averages with the corresponding temporal windows. Background shading marks the magnetic field regime defined by the domain-averaged total magnetic field strength, $\langle|\boldsymbol{B}|\rangle$, with gray for weak-field intervals ($\langle|\boldsymbol{B}|\rangle < 80$ G) and green for strong-field intervals ($\langle|\boldsymbol{B}|\rangle \geqslant 80$ G). The vertical dashed line marks the transition between the two regimes.

For the 10 minute averaging window, the median vortex radius derived from the LCT-derived velocity field remains close to that obtained from the GT velocity field, with values of 1.634 Mm and 1.607 Mm, respectively. As the averaging window increases, the median radius calculated on the LCT-derived velocity field becomes progressively larger, increasing to 1.676, 1.701, and 1.812 Mm for the 20, 30, and 60 minute windows. In contrast, the median radius in the GT velocity field remains 1.607 Mm for all cases. This indicates that longer temporal averaging in the LCT-derived velocities affects the apparent size of the detected vortices, producing larger radii and stronger deviations from the GT measurements. The GT median radius varies smoothly, remaining relatively stable during the weak-field phase and increasing toward later times. The LCT-derived median radius follows the GT velocity field trend most closely for the 10 minute averaging window, while longer temporal averaging produces larger fluctuations and a greater offset between the two curves.

#### 3.1.2. Lifetime Evolution

Vortex lifetimes are estimated by tracking detected vortex centers between consecutive frames using a nearest neighbor association within a fixed spatial threshold. A detection in the following frame is assigned to an existing track if its center lies within two pixels of the previous position. With a pixel scale of 96 km and a cadence of 2 minutes, this corresponds to a maximum displacement speed of approximately $1.6\,\mathrm{km\,s^{-1}}$. This threshold is used only to account for small uncertainties in the detected center position and reasonable frame-to-frame motion. No temporal gaps are allowed, so the lifetime of each vortex is defined as the number of consecutive frames over which the association persists, multiplied by the cadence of 2 minutes. For the representative spatial scale, FWHM $\approx$3.5 Mm, the mean lifetime of vortices detected in the GT velocity field is 2.20 minutes. For the corresponding velocity fields obtained using LCT, the mean lifetime increases with temporal averaging, from 2.49 minutes for the 10 minute window to 3.96, 4.92, and 9.52 minutes for the 20, 30, and 60 minute windows, respectively. This indicates that temporal smoothing biases LCT detections toward longer apparent persistence.

To examine whether the lifetime statistics depend on magnetic field strength, we use the same weak- and strong-field classification defined above, based on the domain-averaged total magnetic field strength $\langle|\boldsymbol{B}|\rangle$ and the threshold $\langle|B|\rangle = 80$ G. Each vortex is assigned to one of these regimes using the value of $\langle|B|\rangle$ at the midpoint of its tracked lifetime. Figure 7 shows the per-frame characteristic lifetime for GT and LCT detections, with the same weak- and strong-field regimes indicated by background shading. The GT velocity field lifetimes remain comparatively stable throughout the time and show only modest variability. In contrast, the lifetimes obtained using LCT are generally larger and exhibit stronger fluctuations, especially for longer temporal averaging windows. The discrepancy is most evident for the 30 and 60 minute windows, in particular during the stronger-field phase. These results show that LCT does not recover lifetime statistics as robustly as it recovers vortex occurrence or radius. It is also clear that the inferred lifetimes become increasingly method-dependent as the effective temporal smoothing increases.

#### 3.1.3. Vorticity Evolution

We next assess whether LCT reproduces the rotational strength of the detected vortices. The rotational strength of each vortex is quantified using the vertical component of the vorticity,

$$\omega_z = \frac{\partial v_y}{\partial x} - \frac{\partial v_x}{\partial y}, \tag{5}$$

computed from both the GT and the LCT-derived velocity fields. For each detected vortex, we calculate the mean absolute vorticity, $|\omega_z|$, over the pixels belonging to each segmented vortex region and then take the median of these per-vortex values in each frame. This provides a representative measure of the vortex vorticity while avoiding cancellation between opposite senses of rotation.

Figure 8 shows the temporal evolution of the median vorticity magnitude for FWHM $\approx$3.5 Mm and temporal averaging windows of 10, 20, 30, and 60 minutes. In the GT velocity field, the median $|\omega_z|$ varies only modestly with time. In contrast, the values obtained by means of LCT show stronger fluctuations and become progressively smaller as the temporal averaging window increases. For FWHM $\approx$3.5 Mm, the mean per-frame median vorticity magnitude in the GT velocity field is $3.64 \times 10^{-4}\ \mathrm{s^{-1}}$. The corresponding values in the LCT-derived fields are $3.72 \times 10^{-4}$, $3.30 \times 10^{-4}$, $3.13 \times 10^{-4}$, and $2.76 \times 10^{-4}\ \mathrm{s^{-1}}$ for the 10, 20, 30, and 60 minute windows, respectively. Thus, while the 10 minute case gives a similar value to the GT velocity field, longer temporal averaging progressively suppresses the recovered vorticity. This shows that LCT is less robust in recovering vortex rotational strength than it is in recovering occurrence or size.

### 3.2. Comparison of Vortex Property Distributions

The time-dependent comparison presented above tests whether LCT reproduces the temporal evolution of the vortex population. In observational applications, however, the GT vortex population is not known, and a strict one-to-one comparison is therefore not possible. We therefore compare the empirical distributions of vortex properties. Following the distinction between continuous and discrete vortex quantities, continuous properties such as radius and area are represented using probability density functions (PDFs), whereas lifetime distributions are represented using probability mass functions (PMFs).

#### 3.2.1. Radius and Area Distributions

The vortex-radius distributions were represented using PDFs. For each averaging window, the GT and LCT radius samples were binned over the fixed interval ($0 \leqslant r \leqslant 3$ Mm). This interval was chosen to cover the radius range of the analyzed vortex samples. The interval was divided into 30 bins, giving a bin width of $\Delta r = 0.1$ Mm. The same bin range and bin width were used for all GT and LCT samples, to ensure a direct comparison between the two distributions. For each radius bin $k$, the empirical probability density was calculated as

$$\hat{f}\,(r_k) = \frac{n_k}{N_r \Delta r}, \tag{6}$$

where $n_k$ is the number of vortices with radii inside bin $k$, $N_r$ is the total number of vortex radii in the corresponding sample, and $\Delta r$ is the bin width. With this normalization, the total area under each histogram is equal to one. Therefore, the plotted quantity represents the empirical probability density of vortex radius, rather than the raw number of detected vortices.

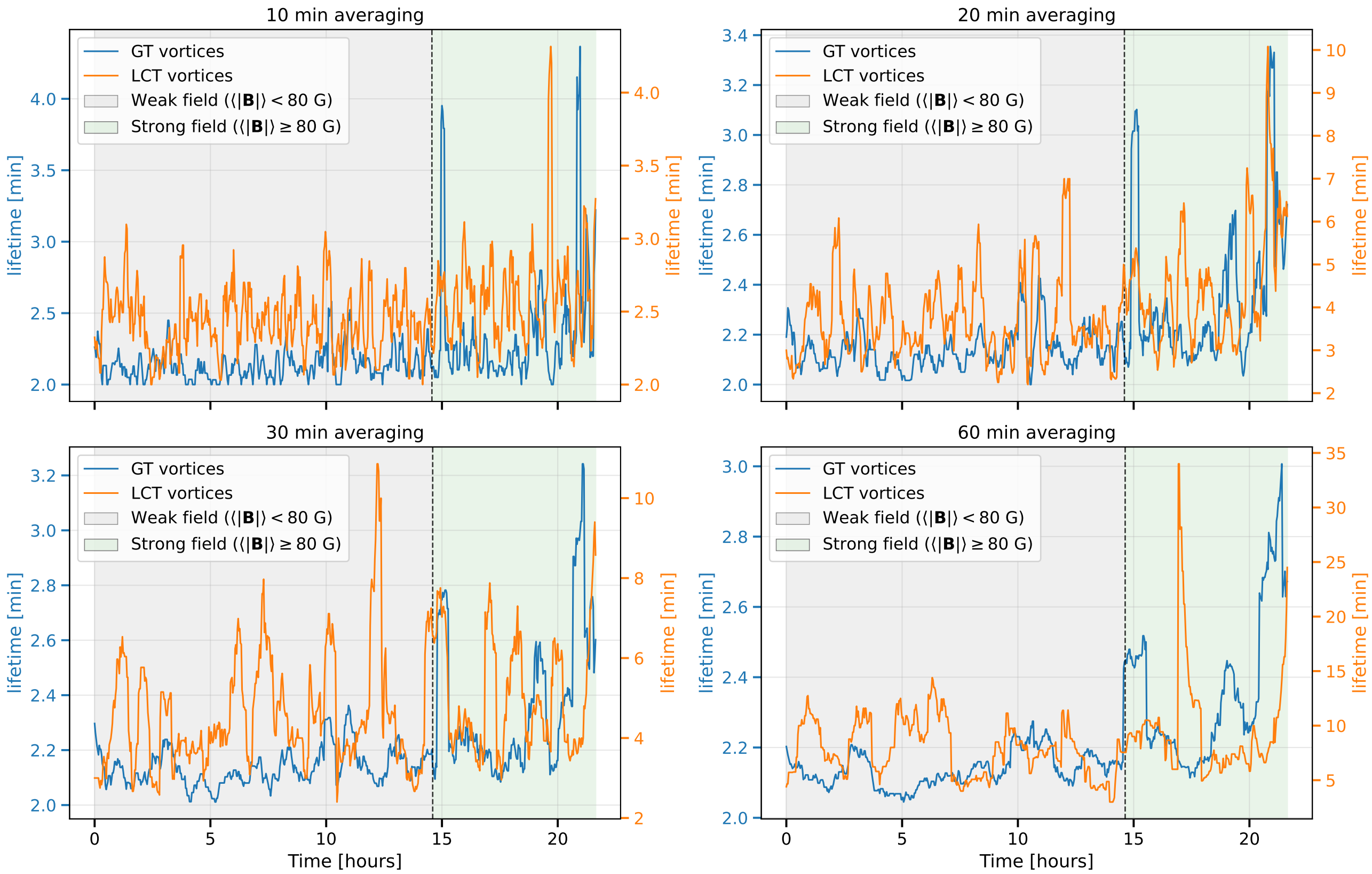

**Figure 7.** Temporal evolution of the mean vortex lifetime in the GT velocity field and in the LCT-derived velocity fields for FWHM ≈3.5 Mm and temporal averaging windows of 10, 20, 30, and 60 minutes. Lifetimes are smoothed using a moving average with the same duration as the corresponding LCT temporal window. Background shading indicates the magnetic field regime based on the domain-averaged total magnetic field strength, $\langle|\boldsymbol{B}|\rangle$: gray regions correspond to weak-field intervals, $\langle|\boldsymbol{B}|\rangle < 80$ G, and green regions correspond to strong-field intervals, $\langle|\boldsymbol{B}|\rangle \geqslant 80$ G. The vertical dashed line marks the transition between the two regimes.

Figure 9 shows the radius distributions for FWHM ≈3.5 Mm and temporal averaging windows of 10, 20, 30, and 60 minutes, separated into weak-field intervals. Across the tested windows, the GT and LCT radius distributions have broadly similar shapes, especially for the 10 minute averaging window. For longer averaging windows, the LCT distributions become broader and show more variability, with a greater contribution from larger vortex radii. This is consistent with the time-evolution analysis, where longer temporal averaging produced larger LCT median radii and a greater deviation from the GT velocity field measurements.

To assess whether this behavior depends on the chosen LCT parameters, the same distribution-based analysis was repeated for all tested combinations of FWHM and temporal averaging windows. The results are summarized in Table 3, where the Spearman correlation coefficients quantify the agreement between the GT and LCT distributions for each physical quantity. The highlighted values indicate the parameter combinations that provide the strongest agreement between the GT and LCT-derived vortex populations. For both radius and area, the correlations generally remain high across most parameter combinations, although the agreement decreases for some longer averaging windows, in particular in the strong-field phase. In contrast, lifetime and vorticity show lower and more variable correlations. This shows that LCT recovers the statistical size properties of vortices more reliably than their temporal or rotational properties.

### *3.2.2. Lifetime Distributions*

The vortex-lifetime distributions were represented using PMFs. For lifetime bin $k$, the probability mass was calculated as

$$p_k = \frac{n_k}{N_t}, \tag{7}$$

where $n_k$ is the number of vortices with lifetimes in bin $k$, and $N_t$ is the total number of vortex tracks in the corresponding sample. This normalization ensures that the sum of the probabilities over all lifetime bins is equal to one.

Figure 10 shows probability mass distributions for the same four temporal averaging windows used earlier, separated into weak- and strong-field regimes. Lifetimes measured in the GT velocity field are strongly concentrated at short durations, with a peak of a few minutes and a rapidly decaying tail. The LCT-derived lifetimes are broader and contain more long-lived detections. This excess of long lifetimes becomes more pronounced as the temporal averaging window increases, consistent with temporal smoothing suppressing short-lived variability and merging intermittent detections into longer-lived tracks. The effect is present in both magnetic regimes, although it is stronger for the 30 and 60 minute windows.

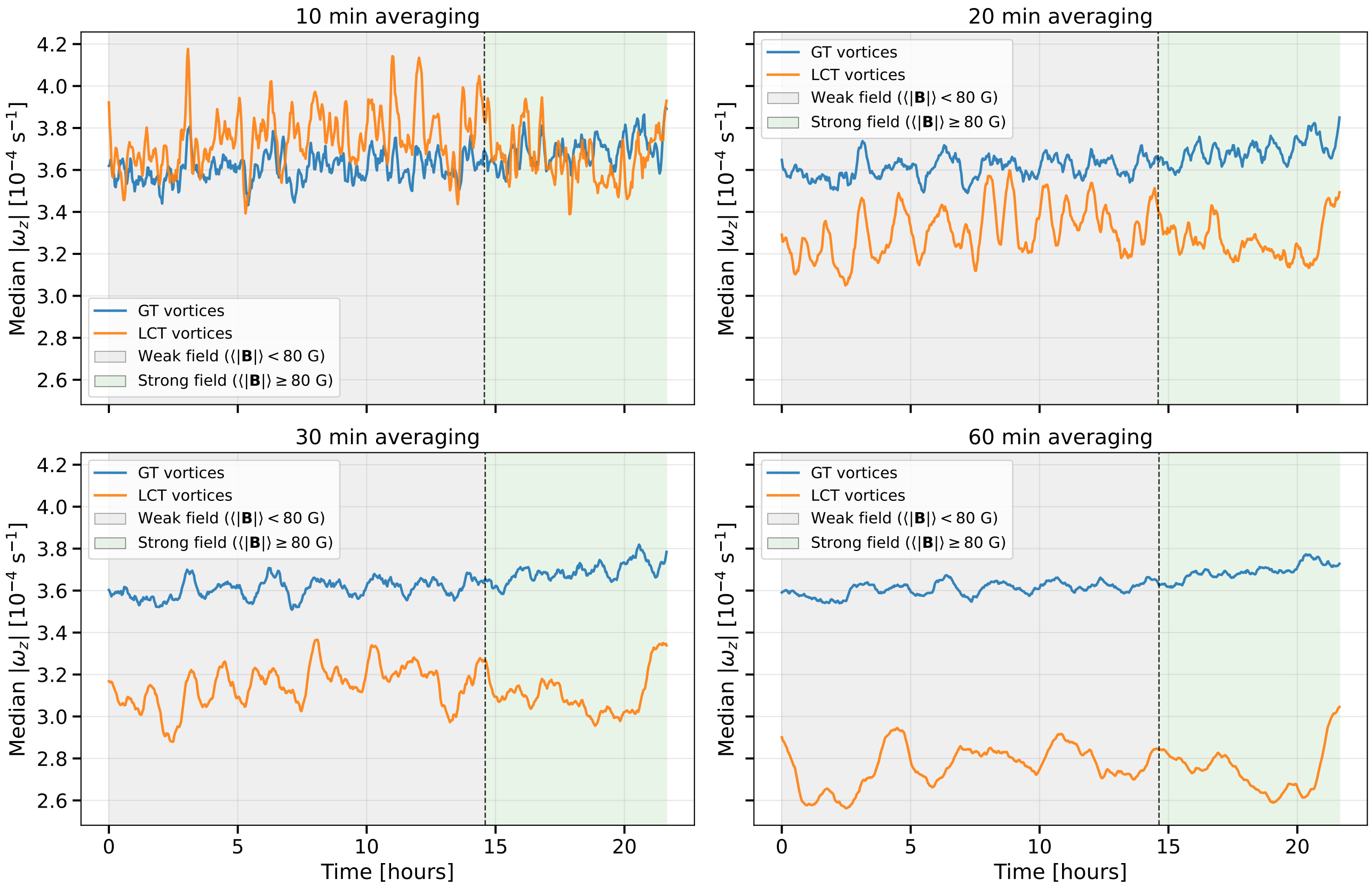

**Figure 8.** Temporal evolution of the median vortex vorticity magnitude, median, $|\omega_z|$, for FWHM ≈3.5 Mm in the GT and LCT-derived velocity fields, for temporal averaging windows of 10, 20, 30, and 60 minutes. The curves are smoothed using centered moving averages of the corresponding duration. Background shading marks weak-field ($\langle|\boldsymbol{B}|\rangle < 80$ G) and strong-field ($\langle|\boldsymbol{B}|\rangle \geqslant 80$ G) intervals. The vertical dashed line marks the transition between the two regimes.

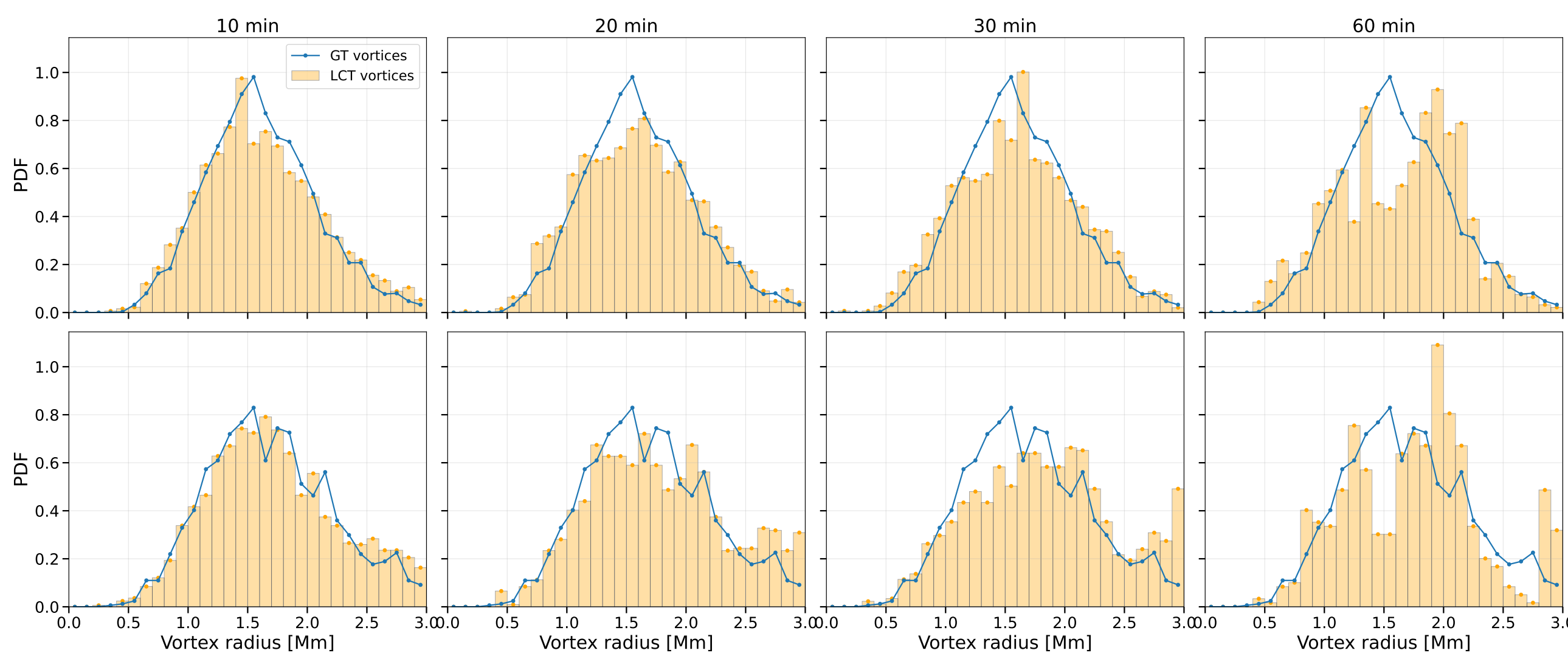

**Figure 9.** Vortex-radius distributions for FWHM ≈3.5 Mm and temporal averaging windows of 10, 20, 30, and 60 minutes, separated into weak-field intervals (top row; $\langle|\boldsymbol{B}|\rangle < 80$ G) and strong-field intervals (bottom row; $\langle|\boldsymbol{B}|\rangle \geqslant 80$ G). The blue curves show the density-normalized radius distributions for vortices detected in the GT velocity field, while the orange histograms show the corresponding distributions for vortices detected in the LCT-derived velocity field. All distributions are normalized to unit area.

The behavior seen in Figure 10 is supported by the full parameter scan summarized in Table 3. Compared with radius and area, the lifetime correlations are lower and more sensitive to both magnetic regime and temporal averaging window. The strongest overall agreement is obtained for FWHM ≈3.5 Mm with 10 minute averaging, but even in this case, the correlation is much weaker in the strong-field intervals than in the weak-field intervals. This indicates that lifetime is less robustly recovered by LCT than the size-related quantities. Long lifetimes inferred from the velocity fields obtained using LCT are strongly affected by the temporal averaging and tracking configuration, and probably represent an upper estimate of vortex persistence.

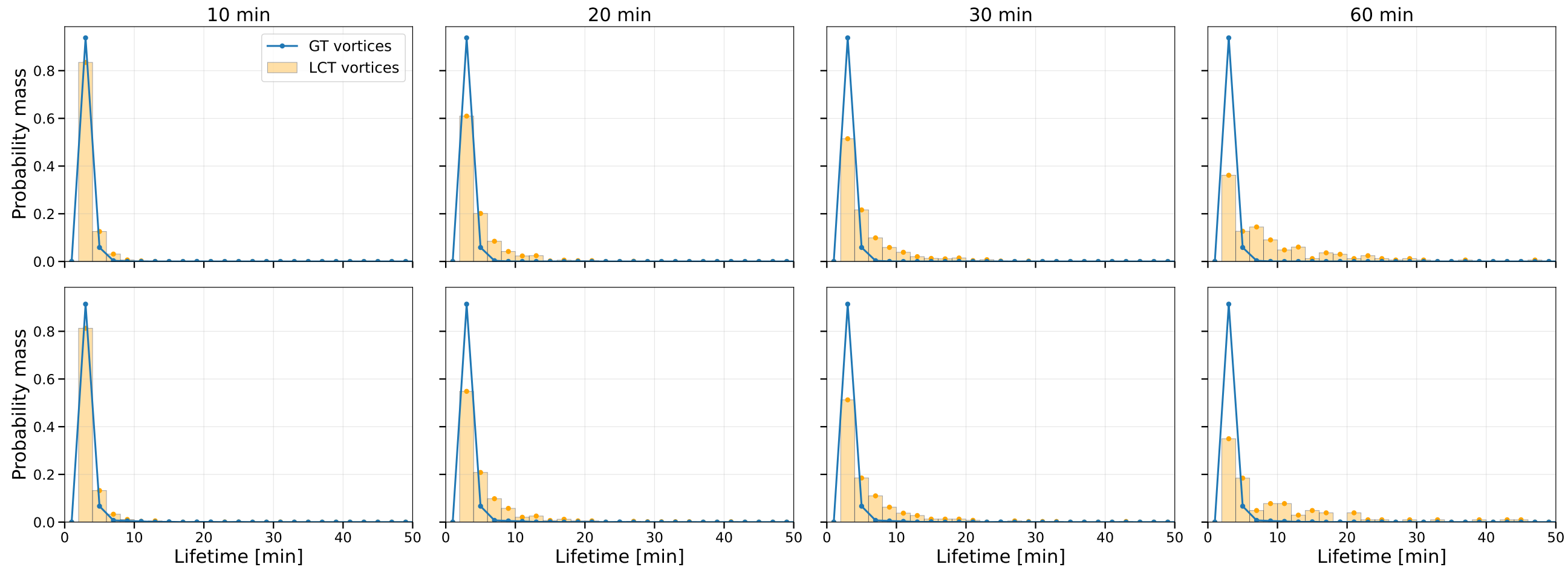

**Figure 10.** Lifetime distributions for vortices detected in the GT and LCT-derived velocity fields for FWHM ≈3.5 and temporal averaging windows of 10, 20, 30, and 60 minutes. The upper and lower rows show weak-field ($\langle|\boldsymbol{B}|\rangle < 80$ G) and strong-field ($\langle|\boldsymbol{B}|\rangle \geqslant 80$ G) intervals, respectively. Blue curves show the GT velocity field detections, and orange bars the LCT-derived detections. All distributions are normalized to unit area.

#### 3.2.3. Vorticity Distributions

We also compare the distributions of vortex vorticity magnitude, using the mean value of $|\omega_z|$ within each segmented vortex region. Unlike radius and area, vorticity depends on spatial derivatives of the velocity field and is, therefore, expected to be more sensitive to smoothing and to differences between the GT and LCT-derived flows.

The full parameter scan summarized in Table 3 shows that the vorticity distributions have lower and more parameter-dependent correlations than the radius and area distributions. The correlations remain relatively high for several parameter combinations, in particular for smaller spatial smoothing scales and shorter averaging windows, but the agreement is less consistent than for size-related quantities. This reduced agreement is consistent with the temporal analysis presented in Section 3.1.3, which showed that temporal averaging progressively suppresses the recovered vorticity. Therefore, vorticity estimates derived from LCT are strongly influenced by the temporal averaging window, in particular at longer averaging times, where the rotational signal is increasingly smoothed.

Table 3 summarizes the corresponding Spearman correlation coefficients for the weak- and strong-field regimes. Our results show that vorticity is more sensitive than radius and area to temporal averaging and to differences between the GT and LCT-derived velocity fields.

## 4. Discussion and Conclusion

We have evaluated how well LCT recovers photospheric vortices in a realistic radiative MHD simulation. Using the R2D2 simulation as a GT, we compared vortices detected from LCT-derived velocity fields with vortices detected from the corresponding spatially smoothed simulation velocity fields.

Since we concentrate on mesogranular scales, the influence of smaller-scale granular variability was reduced by applying a spatial averaging window. The combination of this with different temporal averaging windows allowed us to determine which parameter combinations most successfully recover the GT velocity field and the associated vortex statistics. Vortices were identified using the Γ method applied to both the GT velocities and the LCT-derived flows. This enabled a direct comparison of vortex occurrence, temporal evolution, and statistical distributions as the simulation evolved from a weakly magnetized phase to a more strongly magnetized regime.

LCT captures part of the large-scale temporal variability in vortex occurrence, but does not reproduce the full instantaneous vortex population. In general, the catalogs based on LCT contain fewer vortices than the GT velocity field, and this discrepancy increases with the size of the temporal averaging window. The underestimation is small for the 10 minute averaging window, becomes clearer for the 20 and 30 minute windows, and is strongest for the 60 minute window. The discrepancy also tends to increase during the later, more strongly magnetized phase of the simulation. This is consistent with the smoothing inherent in LCT, which suppresses small-scale velocity gradients, weakens intermittent rotational signatures, and makes short-lived or weak vortices harder to detect. LCT should therefore not be interpreted as recovering the exact vortex population at each time step.

The vortex-count correlations support this interpretation (see Table 2). Spearman's rank correlation between the GT and LCT-derived vortex-count time series remains modest for all tested configurations. The strongest agreement is found for FWHM ≈3.5 Mm with a 30 minute averaging window, for which $\rho \approx 0.53$. This suggests that LCT captures some of the broad temporal variability in vortex occurrence, but does not robustly reproduce the detailed frame-by-frame evolution. The comparison with magnetic field strength shows that both the GT and LCT-derived vortex counts increase during the later, more strongly magnetized phase. However, because the LCT-derived counts remain systematically lower than the GT counts, especially for stronger temporal smoothing, this

**Table 2**
Summary of Vortex-count Statistics for the Tested LCT Configurations

| FWHM (Mm) | $N_{GT}$ | Avg. (minutes) | $N_{LCT}$ | $R_{tot}$ | $\langle R \rangle$ | $\sigma_R$ | Spearman $\rho$ | 95% CI |
|---|---|---|---|---|---|---|---|---|
| 2.5 | 5578 | 10 | 5267 | 0.944 | 1.058 | 0.658 | 0.263 | [0.131, 0.408] |
| 2.5 | 5578 | 20 | 2971 | 0.533 | 0.595 | 0.492 | 0.372 | [0.183, 0.548] |
| 2.5 | 5578 | 30 | 2327 | 0.417 | 0.456 | 0.368 | 0.421 | [0.198, 0.592] |
| 2.5 | 5578 | 60 | 1280 | 0.229 | 0.254 | 0.268 | 0.227 | [−0.241, 0.561] |
| 3.0 | 5605 | 10 | 5544 | 0.989 | 1.096 | 0.556 | 0.287 | [0.162, 0.418] |
| 3.0 | 5605 | 20 | 3173 | 0.566 | 0.620 | 0.373 | 0.437 | [0.265, 0.566] |
| 3.0 | 5605 | 30 | 2566 | 0.458 | 0.500 | 0.354 | 0.509 | [0.305, 0.649] |
| 3.0 | 5605 | 60 | 1515 | 0.270 | 0.297 | 0.256 | 0.306 | [−0.044, 0.567] |
| 3.5 | 5137 | 10 | 5024 | 0.978 | 1.118 | 0.695 | 0.381 | [0.264, 0.497] |
| 3.5 | 5137 | 20 | 3109 | 0.605 | 0.686 | 0.488 | 0.472 | [0.306, 0.583] |
| 3.5 | 5137 | 30 | 2511 | 0.489 | 0.556 | 0.432 | 0.528 | [0.346, 0.637] |
| 3.5 | 5137 | 60 | 1685 | 0.328 | 0.381 | 0.404 | 0.346 | [0.000, 0.590] |
| 4.0 | 4436 | 10 | 4440 | 1.001 | 1.166 | 0.771 | 0.396 | [0.284, 0.493] |
| 4.0 | 4436 | 20 | 2925 | 0.659 | 0.768 | 0.587 | 0.471 | [0.324, 0.596] |
| 4.0 | 4436 | 30 | 2468 | 0.556 | 0.654 | 0.485 | 0.413 | [0.248, 0.551] |
| 4.0 | 4436 | 60 | 1678 | 0.378 | 0.455 | 0.409 | 0.240 | [−0.049, 0.456] |

**Note.** For each spatial window (FWHM) and temporal averaging time, the table lists the integrated vortex counts in the GT and LCT-derived velocity fields, $N_{GT} = \sum_t N_{GT}(t)$ and $N_{LCT} = \sum_t N_{LCT}(t)$, together with the integrated count ratio, $R_{tot} = N_{LCT}/N_{GT}$. It also gives the temporal mean and standard deviation of the per-frame count ratio, $R(t) = N_{LCT}(t)/N_{GT}(t)$. Spearman's rank correlation coefficient, $\rho$, is computed from the smoothed paired time series $N_{GT}(t)$ and $N_{LCT}(t)$ using the same temporal window as the corresponding LCT averaging. The 95% confidence interval is estimated with a block bootstrap.

apparent dependence on the magnetic field must be interpreted together with the recovery limitations of the method. A separate comparison using the local magnetic field strength within individual GT vortex cores showed no systematic dependence of LCT recovery on core-field strength.

The comparison based on distributions provides a complementary perspective. While the time-dependent analysis tests whether LCT follows the same temporal evolution, the distribution-based analysis tests whether LCT recovers the statistical properties of the vortex population. This distinction is important because, in observations, the GT vortex population is unknown, and a strict one-to-one validation is not possible. Our results show that LCT performs best for size-related quantities. In particular, the radius and area distributions generally show high Spearman correlations between the GT and LCT-derived probability-density distributions across most FWHM values, temporal averaging windows, and magnetic field regimes (see Table 3). The agreement decreases for some longer averaging windows, especially in the strong-field phase, but it remains stronger than for lifetime and vorticity. This indicates that, even when individual vortices are not recovered at the same locations and times, LCT can still reproduce the statistical size distribution of the vortex population.

The radius and area results illustrate this most clearly. The time-dependent analysis shows that the 10 min averaging window gives the closest agreement between the LCT-derived and GT median radii. For FWHM ≈3.5 Mm, the median LCT radius is 1.634 Mm for the 10 min window, compared with 1.607 Mm in the GT velocity field. As the temporal averaging window increases, the LCT median radius becomes larger, reaching 1.676, 1.701, and 1.812 Mm for the 20, 30, and 60 minute windows, respectively. This indicates that longer temporal averaging affects the apparent vortex size and can lead to larger deviations from the GT radius. However, the distribution-based comparison shows that the radius and area distributions remain broadly similar across the tested temporal averaging windows, and magnetic field regimes (see Figure 9). Thus, even though LCT does not recover the full instantaneous vortex population, it still retains useful statistical information about vortex size.

Lifetime and vorticity are recovered less robustly than the size-related quantities. For the representative case with FWHM ≈3.5 Mm, the LCT-derived mean lifetime increases from 2.49 minutes for the 10 minute window to 9.52 minutes for the 60 minute window, compared with 2.20 minutes in the GT velocity field. This shows that temporal smoothing biases LCT detections toward longer apparent persistence, likely by suppressing short-lived variability and linking intermittent detections into longer-lived tracks. The lifetime distributions confirm this interpretation: the GT lifetimes are concentrated at short durations, whereas the LCT-derived distributions are broader and develop more extended tails, especially for the 30 and 60 minute windows. Therefore, LCT-derived lifetimes should be regarded as method-dependent estimates rather than direct measurements of the physical persistence of individual vortices.

The vorticity results also show the effect of smoothing. For FWHM ≈3.5 Mm, the mean per-frame median vorticity magnitude in the GT velocity field is $3.64 \times 10^{-4}\,s^{-1}$. The 10 minute LCT case gives a comparable value, $3.72 \times 10^{-4}\,s^{-1}$, but longer temporal averaging progressively reduces the recovered vorticity, reaching $2.76 \times 10^{-4}\,s^{-1}$ for the 60 minute window. This is expected because vorticity depends on spatial derivatives of the velocity field and is therefore sensitive to smoothing, noise, and differences between the GT and the LCT-inferred velocity fields. Similar to this, the vorticity distribution correlations are lower and more parameter-dependent than those for radius and area.

The strong sensitivity of vortex counts to smoothing in our LCT-derived fields is consistent with published FLCT-based validation work. S. S. Turkay et al. (2026) showed that the

**Table 3**
Spearman Correlation Coefficients between the GT and LCT-derived Probability-density Distributions

| Quantity | FWHM [Mm] | 10 min | 20 min | 30 min | 60 min |
|---|---|---|---|---|---|
| Radius | 2.5 | 0.980 – 0.967 | 0.976 – 0.937 | 0.971 – 0.892 | 0.969 – 0.508 |
| Radius | 3.0 | 0.989 – 0.969 | 0.972 – 0.933 | 0.970 – 0.900 | 0.939 – 0.662 |
| Radius | 3.5 | 0.986 – 0.961 | 0.972 – 0.886 | 0.986 – 0.799 | 0.826 – 0.664 |
| Radius | 4.0 | 0.958 – 0.925 | 0.957 – 0.918 | 0.951 – 0.885 | 0.806 – 0.862 |
| Area | 2.5 | 0.970 – 0.961 | 0.985 – 0.957 | 0.941 – 0.891 | 0.968 – 0.735 |
| Area | 3.0 | 0.977 – 0.944 | 0.964 – 0.935 | 0.962 – 0.947 | 0.952 – 0.764 |
| Area | 3.5 | 0.978 – 0.954 | 0.970 – 0.973 | 0.973 – 0.941 | 0.934 – 0.719 |
| Area | 4.0 | 0.981 – 0.911 | 0.943 – 0.892 | 0.954 – 0.911 | 0.915 – 0.655 |
| Lifetime | 2.5 | 0.837 – 0.762 | 0.777 – 0.719 | 0.814 – 0.740 | 0.740 – 0.682 |
| Lifetime | 3.0 | 0.941 – 0.754 | 0.795 – 0.777 | 0.765 – 0.720 | 0.671 – 0.729 |
| Lifetime | 3.5 | 0.956 – 0.819 | 0.893 – 0.598 | 0.846 – 0.572 | 0.721 – 0.646 |
| Lifetime | 4.0 | 0.894 – 0.716 | 0.709 – 0.657 | 0.639 – 0.684 | 0.539 – 0.818 |
| Vorticity | 2.5 | 0.927 – 0.958 | 0.853 – 0.741 | 0.948 – 0.938 | 0.933 – 0.917 |
| Vorticity | 3.0 | 0.868 – 0.870 | 0.910 – 0.879 | 0.914 – 0.872 | 0.933 – 0.910 |
| Vorticity | 3.5 | 0.846 – 0.838 | 0.859 – 0.837 | 0.862 – 0.835 | 0.901 – 0.871 |
| Vorticity | 4.0 | 0.777 – 0.778 | 0.797 – 0.786 | 0.838 – 0.795 | 0.896 – 0.891 |
| [ | | | | | |

**Note.** Each cell gives the weak-field and strong-field values, respectively, in the format weak–strong. The green highlights indicate the selected best performing parameter combination for each physical quantity, based on the overall agreement between the weak- and strong-field distributions.

number of FLCT vortex events decreases rapidly as the Gaussian kernel width increases. They also emphasized that, although the resulting vortex-area and lifetime distributions remain broadly comparable between DKIST observations and synthetic MURaM intensity data, the inferred vortex population is strongly sensitive to the tracking parameters. In our analysis, the intermediate-smoothing configuration (FWHM ≈3.5 Mm with a 20 minute averaging window) gives $\langle \mathcal{R} \rangle \approx 0.69$, which is quantitatively similar to the ratio reported by S. S. Turkay et al. (2026) for the moderately smoothed MURaM subset (6969 FLCT events versus 10,154 vortices detected from the GT velocity field, i.e., ∼0.69). Although our analysis targets larger mesogranular-scale structures and includes temporal averaging, the overall behavior is similar, i.e., stronger effective smoothing reduces the number of detected vortices.

To place our results in the context of previous vortex statistics, we computed the spacetime vortex density as $d = N_v/(A\,T)$, where $N_v$ is the total number of detected vortices, $A$ is the analyzed area, and $T$ is the duration of the time series. For the representative spatial window, FWHM ≈3.5 Mm, using an analyzed domain of $98 \times 98\,\mathrm{Mm}^2$ and a duration of $T = 1320$ minutes, the GT vortex density is $4.05 \times 10^{-4}\,\mathrm{Mm}^{-2}\,\mathrm{minute}^{-1}$. The corresponding LCT-derived densities decrease with increasing temporal averaging, from $3.96 \times 10^{-4}\,\mathrm{Mm}^{-2}\,\mathrm{minute}^{-1}$ for the 10 minute window to $2.45 \times 10^{-4}$, $1.98 \times 10^{-4}$, and $1.33 \times 10^{-4}\,\mathrm{Mm}^{-2}\,\mathrm{minute}^{-1}$ for the 20, 30, and 60 minute windows, respectively. This confirms that stronger temporal averaging progressively reduces the number density of vortices recovered by LCT. These values are much smaller than those reported for small-scale photospheric intensity vortices by I. Giagkiozis et al. (2018), who found $d = 0.84\,\mathrm{Mm}^{-2}\,\mathrm{min}^{-1}$ using an effective field of view of approximately $39.9 \times 39.9\,\mathrm{Mm}^2$. Similarly, S. S. Turkay et al. (2026) reported substantially larger instantaneous vortex number densities for small-scale MURaM and FLCT-derived vortices. They also found that the spatial overlap between vortices identified in the velocity field derived from the MURaM intensity data with use of FLCT and GT vortices was only 0.4%–19.7%, depending on the FLCT kernel size. This is comparable to the low LCT–GT overlap fraction found in the present study, 3.54%–17.88%, although the two values should not be compared directly, because the simulations, vortex scales, velocity-inference methods, smoothing parameters, and overlap criteria are different. These comparisons are therefore qualitative and scale-dependent, since the studies differ in domain size, duration, resolution, detection method, and characteristic vortex size. The smaller densities found here are expected because our analysis targets larger-scale vortical structures over a larger field of view rather than small intergranular vortices.

Several improvements could strengthen future applications of vortex identification using the Γ method applied to LCT-derived velocity fields. For example, the tracking stage could be extended to include multiple spatial scales, and the recovered flows could be compared with those obtained from alternative plasma flow estimation methods such as DAVE (P. W. Schuck 2006). A useful complementary test would be to examine whether the inferred motions are consistent with the observed displacement of magnetic field elements, for example, using magnetic feature tracking tools such as SoFT

(M. Berretti et al. 2025). Such comparisons would provide a more robust assessment of both the recovered velocity field and the resulting vortex statistics. The method could then be validated across a wider range of surface phenomena and applied to long-duration observations, such as SDO/HMI, to study coherent vortical activity over the solar cycle. These extensions would be valuable, but they would also require careful and uniform preprocessing, as well as strict control of the analysis parameters, to avoid introducing artificial trends. Since these modifications need to be thoroughly validated for broader applications, they fall outside the scope of this study.

Since LCT was evaluated here using simulated intensity data, its application to observation data may require additional considerations. Instrumental effects such as image degradation, stray light, and reduced contrast can bias the inferred velocity field. Differences in cadence, spatial resolution, and oscillatory intensity fluctuations can also affect the recovered vortex statistics. Appropriate preprocessing, including PSF correction, intensity normalization, and filtering of oscillatory signals, is therefore essential when applying LCT to observational time series (e.g., K. Yeo et al. 2014; B. Tremblay et al. 2018; I. Giagkiozis et al. 2018). Consistent preprocessing and resolution matching are particularly important when comparing different datasets.

The reliability of LCT also depends on how well the intensity data satisfy the assumptions of the method. It performs best when the image sequence contains coherent, trackable features at the chosen scale and when inter-frame displacements remain small relative to the tracking window. If these conditions are not met, the recovered velocities can deviate from the underlying flow, and the inferred vortex statistics may become biased. Because LCT acts as a scale-dependent filter set by the adopted FWHM and temporal averaging, vortex catalogs are only directly comparable when tracking parameters and effective resolution are matched. In our tests, increasing the temporal averaging from 10 to 60 minutes strongly affected both the completeness and the persistence of the recovered vortices.

Our results show that applying the $\Gamma$ method to LCT-derived velocity fields is best interpreted as a statistical diagnostic rather than as a tool for precise one-to-one mesogranular vortex identification. LCT captures part of the broad temporal variability in vortex occurrence, but it underestimates the absolute number of vortices and does not recover the full instantaneous vortex population. The most robustly recovered quantities are the statistical size distributions, in particular radius and area, whereas lifetime and vorticity are more sensitive to the adopted FWHM and temporal averaging window. We therefore conclude that applying the $\Gamma$ method to LCT-derived velocity fields is suitable for studying smoothed, ensemble-level vortical behavior in long intensity sequences, provided that the method is calibrated against simulations and that the smoothing parameters are included in the physical interpretation.

## Acknowledgments

M.A. acknowledges the University of Hafr Al Batin and the Ministry of Education in the Kingdom of Saudi Arabia for their financial support. V.F. and S.S.A.S. are grateful to the Science and Technology Facilities Council (STFC) grants ST/V000977/1, ST/Y001532/1. V.F. is grateful to the STFC grant UKRI1165. V.F., S.S.A.S., I.B., and G.V. are thankful to The Royal Society, International Exchanges Scheme, collaboration with Greece (IES/R1/221095) and The Royal Society, IEC/R3/233017—International Exchanges 2023 Cost Share (NSTC), collaboration with Taiwan. S.S.A.S. and V.F. would like to thank the International Space Science Institute (ISSI) in Bern, Switzerland, for the hospitality provided to the members of the "Opening new avenues in identifying coherent structures and transport barriers in the magnetized solar plasma" team (ISSI Team 607).

## ORCID iDs

Mashael Aldhafeeri https://orcid.org/0009-0000-9098-8240
Suzana S. A. Silva https://orcid.org/0000-0001-5414-0197
Istvan Ballai https://orcid.org/0000-0002-3066-7653
Gary Verth https://orcid.org/0000-0002-9546-2368
Haruhisa Iijima https://orcid.org/0000-0002-1007-181X
Hideyuki Hotta https://orcid.org/0000-0002-6312-7944
Viktor Fedun https://orcid.org/0000-0002-0893-7346